# Dissipative Collisions of Impinging Liquid Jets Having Uniform Velocity Profiles

Robert J. Demyanovich[1]
*RJD Technologies, Inc., Seattle, WA 98146, USA*

**The dissipation of energy in the impingement zone from the collision of impinging, free, equal jets of liquids was investigated by comparison with studies on the energy dissipation from the collision of impinging, free, equal sheets of liquids. Loss of energy was studied in terms of the coefficient of restitution ($COR$) of the collision. With few exceptions, previous analytical studies have assumed that there is no loss of energy resulting from the collision of the jets ($COR$ = 1), and for jets with uniform velocity profiles, the sheet velocity (velocity after the collision) is equal to the jet velocity (velocity before the collision). In this study, mass and momentum balances of impinging jets with uniform velocity profiles are revised to include the impact of non-equal velocities ($COR$ < 1). After development of the applicable theory, the $COR$ for impinging jets is calculated from available data in the literature on jet and sheet velocities as well as the location of the stagnation point in the impingement zone. Simple empirical correlations were developed for 1) the $COR$ of impinging jets as a function of impingement angle and 2) the relationship of the location of the stagnation point to the $COR$. A theoretical equation was also derived for the impact of dissipative collisions on the thickness distribution in the liquid sheet.**

## Nomenclature

| | | |
|---|---|---|
| $b_{HP}$ | = | for a 100% elastic collision of impinging jets, distance from center of ellipse to separation point in cross section through the jet or to the stagnation point in the plane of symmetry of the impingement zone [8] |
| $d_j$ | = | jet diameter |
| $d_o$ | = | orifice diameter |
| $h$ | = | thickness of the sheet produced by impinging jets |

---

[1] President, RJD Technologies, Inc.

| | | |
|---|---|---|
| $h_\pi$ | = | thickness of the sheet produced by impinging jets at the azimuthal angle equal to $\pi$ radians |
| $L$ | = | length of free impinging jet from tube exit to impingement zone |
| $\dot{m}$ | = | mass flowrate of liquid in a single sheet |
| $\dot{m}_b$ | = | mass flowrate of liquid in the backward direction of the mixed sheet formed from impinging sheets |
| $\dot{m}_f$ | = | mass flowrate of liquid in the forward direction of the mixed sheet formed from impinging sheets |
| $M$ | = | separation point in the cross-section through the jet |
| $p$ | = | for a partially elastic collision, distance from the separation point to a point on the ellipse of the cross-section through jet |
| $p_\pi$ | = | distance, $p$, at an azimuthal angle equal to $\pi$ radians |
| $q$ | = | for a 100% elastic collision, distance from the separation point to a point on the ellipse of the cross-section through jet |
| $\Delta P$ | = | injector pressure drop |
| $r$ | = | distance from the stagnation point to the sheet thickness, h |
| $r_\pi$ | = | radial line distance, $r$, at an azimuthal angle equal to $\pi$ radians |
| $R$ | = | jet radius |
| $R_i$ | = | distance from the origin of the single sheets to the impingement zone |
| $R_B$ | = | distance from the origin of the single sheets to the rim of the backward sheet |
| $s_i$ | = | thickness of the single sheets at impingement |
| $S$ | = | stagnation point |
| $S_{HP}$ | = | stagnation point for a 100% elastic collision |
| $u$ | = | velocity of single sheets |
| $u_c$ | = | for impinging sheets, single-sheet critical velocity that results in a partially elastic collision |
| $v_j$ | = | uniform velocity of jets |
| $v_m$ | = | velocity of mixed sheet formed by impinging sheets |
| $v_s$ | = | velocity of sheet formed by impinging jets |
| $v_{s,c}$ | = | velocity of sheet where the Weber number of the sheet is equal to 1. |
| $v_{TC}$ | = | Taylor-Culick velocity |
| $v_x$ | = | component velocity of the single-sheet velocity or jet velocity in the x direction |
| $v_y$ | = | component velocity of the single-sheet velocity or jet velocity in the y direction |
| $w$ | = | distance from the center of the ellipse to the separation point in the cross section through the jet or the stagnation point in the plane of symmetry of the impingement zone of impinging jets |
| $We_{bs}$ | = | for impinging sheets, Weber number at the rim of the backward mixed sheet |
| $We_{i,c}$ | = | for impinging sheets, Weber number at the impingement zone and at the critical single-sheet velocity for a partially elastic collision |
| $We_{j,c}$ | = | for impinging jets, Weber number at the critical velocity for a partially elastic collision, $\rho d_j v_{j,c}^2/\sigma$ |

*Greek Letters*

| | | |
|---|---|---|
| $\beta$ | = | parameter for the thickness distribution correlation by Ibrahim and Przekwas [6] |
| $\kappa$ | = | azimuthal angle of a radial line measured from the separation point, M, or the stagnation point, S |
| $\varphi$ | = | for a 100% elastic collision, azimuthal angle of a radial line measured from the separation point, M, or the stagnation point, $S_{HP}$ |
| $\rho$ | = | liquid density |
| $\Psi$ | = | coefficient of effective velocity defined by Tokuoka and Sato[18], which is the ratio of the sheet velocity to the jet velocity |
| $\Psi_{180^o}$ | = | coefficient of effective velocity at $2\theta = 180°$ [18] |
| $\sigma$ | = | surface tension |
| $\theta$ | = | half angle of impingement |

*Acronyms*

| | | |
|---|---|---|
| *COR* | = | coefficient of restitution of the collision of impinging streams |
| $COR_j$ | = | coefficient of restitution of the collision of impinging jets |
| $COR_{j,c}$ | = | *COR* of the collision of impinging jets at the critical velocity for a partially elastic collision |
| $COR_s$ | = | coefficient of restitution of the collision of impinging sheets |
| $COR_{s,c}$ | = | *COR* of the collision of impinging sheets at the critical velocity for a partially elastic collision |

## I. Introduction

The impingement of obliquely oriented jets is frequently used for the injection of liquid rocket propellants. Although the combustion process occurs in the gas phase, the impingement of liquid jets provides good distribution and atomization of the liquid propellants prior to combustion [1]. Further, impinging jet injectors are relatively simple to fabricate.

In rocket engines, impinging-jet injectors produce identical, cylindrical jets which are caused to impinge upon one another at an angle (2θ) typically less than 90º. If the momentum of both jets is equal, the impingement produces a liquid sheet that is normal to the plane of impingement. Since the sheet expands, it becomes thinner with distance from the impingement zone. The sheet disintegrates into ligaments, which then contract due to surface tension ultimately breaking up into droplets.

The rate-controlling step of the gas-phase combustion is, in most cases, vaporization of the liquid propellants. The rate of vaporization is strongly affected by the atomization of the liquid propellants [2,3]. According to Priem and Heidmann [2], the vaporization rate is inversely proportional to the mass median droplet radius to the 1.45 power.

Drop sizes and distributions resulting from impinging-jet injectors are, in turn, dependent upon the characteristics of the formed liquid sheet [4,5]. In general, the characteristics of both the liquid sheet and the resulting spray depend upon the impingement angle, the jet diameter, the jet velocity, whether the pre-impinging jets are laminar or turbulent, the length of the jets to the impingement zone, the physical properties of the liquid, and the characteristics of the gas phase (such as absolute pressure) [4].

Much research has been devoted to understanding the breakup characteristics of the liquid sheet formed by the impingement of two equal jets. Since the resulting drop size is a function of the sheet thickness at breakup, previous analytical studies have included the modeling of sheet thickness with distance from the impingement zone [6–9]. These studies investigated the thickness parameter or how the sheet thickness normalized by the jet radius varies with both radial distance from the impingement zone and azimuthal angle around the sheet. In addition, studies have focused on the sheet breakup mechanisms [10,11] and the sheet breakup length [12–14]. Velocities of the liquid sheet [4,10,15] and drop sizes and drop-size distributions have also been investigated [5,10,12–14,16].

Analyses involving mass and momentum balances of the colliding jets have almost universally assumed that no energy is lost as a result of the collision. For jets with uniform velocity profiles, this assumption results in the velocity of the formed liquid sheet being set equal to the velocity of the pre-impinging jets.

Exceptions include Heidmann et al., [17] who assumed that the velocity in the sheet was equal to the axial component velocity of the jet ($v_j$cosθ); however, this assumption was not based on the results of any experimental measurements. Another exception is Tokuoka and Sato [18], who state that the velocity of the liquid sheet formed by the impingement of two directly opposed jets (2θ = 180º) was experimentally measured as 80% of the jet velocity. They propose a method for estimating the sheet velocity at oblique angles of impingement; but this method does not appear to be based on the results of additional experimental measurements or any process simulations. In a subsequent analysis of the characteristics of the formed liquid sheet, Inamura and Shirota [19] incorporated the work of Tokuoka and Sato [18] and assumed that the sheet velocity was equal to 80% of the jet velocity at all impingement angles. A third exception is a computational fluid dynamics (CFD) study on impinging jets by Chen, et al. [20]. As part of that investigation, it was found that the location of the stagnation point in the impingement zone shifts if the collision is dissipative.

Several studies investigating the spray velocity have noted that the resulting velocity of the sheet/droplets is closer to the jet velocity, $v_j$, than to $v_j\cos\theta$ [4,12,21]. Choo and Kang [4] state that "*The magnitude of the maximum velocity of ligaments observed around the axis of the spray is a little lower than the corresponding jet velocity but a little higher than the value $v_j\cos\theta$ at all jet velocities. This is consistent with the previously held belief that the spray velocity is usually equal to the jet velocity.*"

Research on impinging free equal sheets of liquids has identified three types of collisions for impinging sheets [22]: 1) "100% elastic" where no energy is lost due to the collision, 2) "inelastic" where energy is dissipated in the impingement zone [23] and the mixed or combined-sheet velocity is theoretically equal to the axial component velocity of the single sheets ($u\cos\theta$), and 3) "partially elastic" where the velocity in the resulting mixed sheet is between that for a 100% elastic collision and an inelastic collision. A major portion of the present work is an analysis of the applicability of partially elastic collisions of impinging sheets to impinging jets. Partially elastic collisions may be responsible for the observations that the spray velocity is in between the velocity of the jets ($v_j$) and the axial component velocity of the jets ($v_j\cos\theta$) [4,12,21].

The distribution of liquid around the liquid sheet is a function of the impingement angle ($2\theta$) of the jets. At $2\theta < 180º$, more liquid flows in the forward part of the sheet than in the backward part of the sheet with this disparity increasing as $2\theta$ is decreased. Further, in liquid rocket engines, the amount of flow in the backward direction, towards the impinging-jet injector, can result in injector-face burnout and is one of the reasons why the impingement angle is typically set at less than 90º [24]. The effect of dissipative collisions on this thickness distribution is discussed in Section IV.

## II. Coefficient of Restitution of the Collision

### A. Impinging Sheets

The energy dissipated from the collision of impinging jets was analyzed using the same methodology as in the study on the energy dissipation resulting from the collision of impinging sheets [22]. A perspective view of the impingement of equal liquid streams is shown in Fig. 1. Fig. 1a illustrates the impingement of two equal jets while Fig. 1b depicts the impingement of two, equal single sheets. For impinging jets, the collision of the jets produces a liquid sheet, whereas for impinging sheets, the collision of the single sheets produces a mixture of the liquids, which is also in the form of a continuous sheet (mixed sheet). The plane of the liquid sheet after impingement is determined by the momentum of the jets or the single sheets. When the jets or single sheets are equal, the liquid sheet or mixed liquid sheet is formed in a plane that is a bisection of the impingement angle. At equal flowrates, the thickness of the single sheets at impingement is typically 20 to 30 times less than the diameter of the equivalent jets.

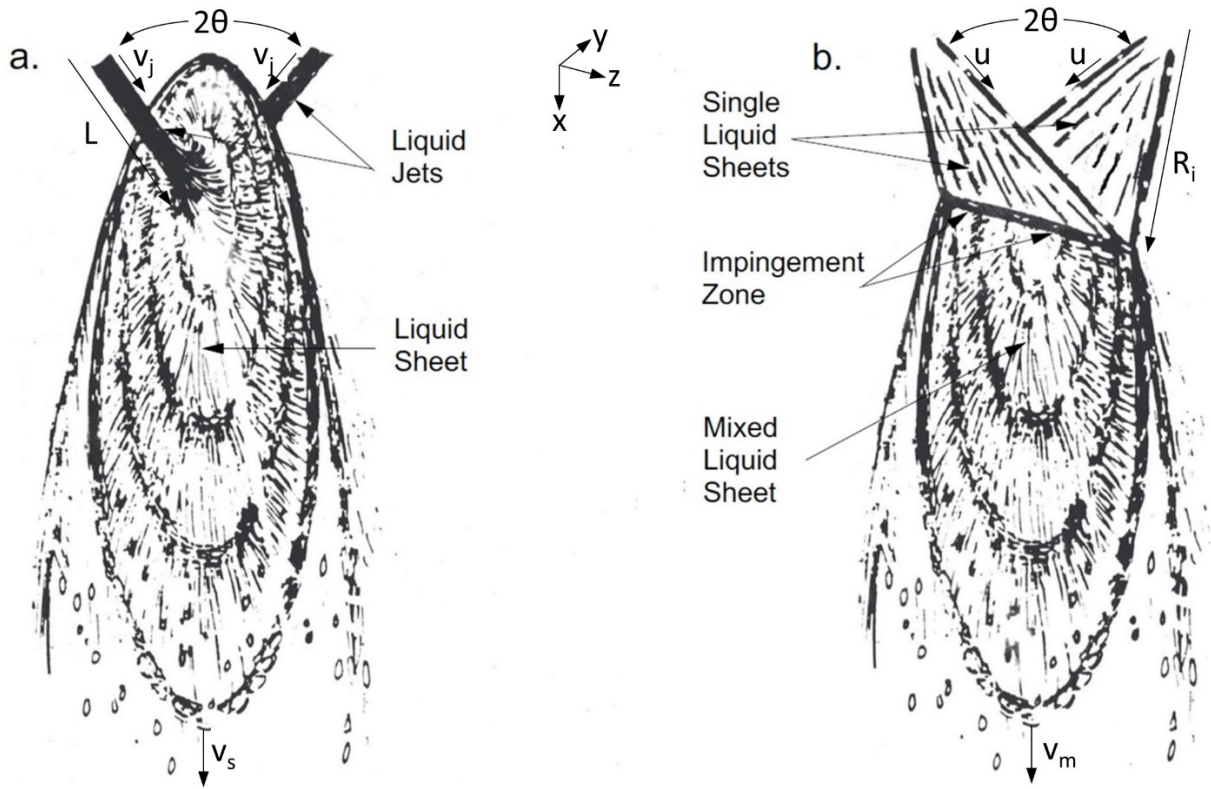

**Fig. 1. Impingement of two equal streams of liquid at an angle of 2θ. Fig. 1a is a perspective view of the impingement of two, equal cylindrical jets traveling at a velocity of $v_j$ producing a liquid sheet with a velocity of $v_s$. Fig. 1b is a perspective view of the impingement of two, equal thin sheets traveling at a velocity of $u$ producing a mixed sheet with a velocity of $v_m$. L – length of free impinging jet from tube exit to impingement zone; $R_i$ – radial distance from the origin of the single sheets to the impingement zone.**

The coefficient of restitution ($COR$) of the collision of impinging sheets has been investigated for half angles of impingement (θ) between 20° and 60°, single-sheet thickness at impingement ($s_i$) between 30 and 160 μm, single-sheet velocities ($u$) ranging from 3 to 12 m/s and for various liquids including water, 90% ethanol, 32% aqueous glycerine and skim milk [22].

For impinging sheets, the velocity of single sheets is constant everywhere in the liquid sheet [22,25–28]. The velocity in the mixed sheet (not the impingement zone) was also found to be constant everywhere within the mixed sheet [22,23].

Since the mass flow rates are equal, single-sheet velocities are constant and equal, and the mixed-sheet velocity is constant, the mass and momentum balances for the collision of two equal sheets can be simply expressed as:

$$2\dot{m} = \dot{m}_f + \dot{m}_b \tag{1}$$

and

$$\dot{m}v_x + \dot{m}v_x = \dot{m}_f v_m + \dot{m}_b v_m \tag{2}$$

$$\dot{m}v_y - \dot{m}v_y = 0 \tag{3}$$

where $\dot{m}$ is the mass flow rate of each single sheet, $\dot{m}_f$ is the mass flow rate of the mixed sheet in the forward direction, $\dot{m}_b$ is the mass flow rate of liquid in the backward direction, $v_m$ is the velocity of liquid in the mixed sheet and $v_x$ and $v_y$ are the component velocities of the single-sheet velocity ($u$).

The coefficient of restitution ($COR$) of the collision for impinging sheets ($COR_s$) is defined as,

$$COR_s = \frac{v_m}{u} \tag{4}$$

Further analysis [22] leads to the following,

$$\frac{\dot{m}_f}{\dot{m}} = \frac{COR_s + \cos\theta}{COR_s} \tag{5}$$

$$\frac{\dot{m}_b}{\dot{m}} = \frac{COR_s - \cos\theta}{COR_s} \tag{6}$$

and

$$\frac{\dot{m}_f}{\dot{m}_b} = \frac{COR_s + \cos\theta}{COR_s - \cos\theta} \tag{7}$$

Thus, the ratio of flows is a function of the $COR$ of the collision. If the collision is 100% elastic ($v_m = u$), then $COR_s = 1$; if the collision is inelastic, then $COR_s = \cos\theta$; and if partially elastic, $\cos\theta < COR_s < 1$. At conditions studied to date, experiments have not measured values of $COR_s$ that indicate a 100% elastic collision.

Figure 2 contains super-macro photographs focused on the impingement zone of impinging sheets. Measured values of the $COR$ for collisions where liquid is ejected from the backward mixed sheet (left photograph) indicate partially elastic collisions ($\cos\theta < COR_s < 1$). The measured $COR$ for collisions where liquid is not ejected from the backward mixed sheet (right photograph) indicate an inelastic collision ($COR_s \approx \cos\theta$). If the collision is inelastic, Eq. 6 indicates that there is no backflow (although a backward mixed sheet is visible, liquid is not ejected from the backward mixed sheet).

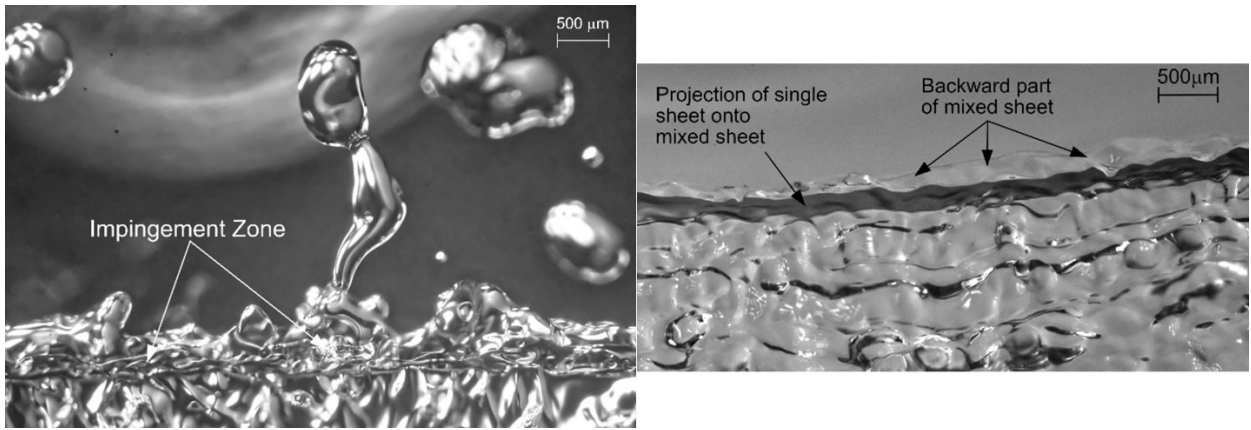

**Fig. 2. Left photograph (Fig. 12 in Demyanovich [31]) – four μs flash photograph of the front view of a <u>partially elastic collision</u> of two equal sheets of 40% aqueous glycerin at an impingement angle (2θ) of 110º, a pressure drop (ΔP) of 20 kPa, an equivalent jet orifice diameter (d$_o$) of 0.19 mm, a distance to impingement (R$_i$) of 2 cm and a calculated sheet thickness at impingement (s$_i$) of 92 μm. Right photograph (Fig. 2 in Demyanovich [22]) – four μs flash photograph of the front view of an <u>inelastic collision</u> of two equal sheets of water at 2θ = 90º, ΔP = 25 kPa, d$_o$ = 0.19 mm, R$_i$ = 2 cm and s$_i$ = 87 μm.**

The rotating ligaments in the left photograph verify that turbulence is created in the impingement zone of impinging sheets resulting from the energy released by the collision and subsequently dissipated. This turbulence has been shown to yield rapid micromixing of the impinging liquids [29]. The turbulence, however, dissipates within approximately one turnover time of the largest eddies in the impingement zone [23] and often is not fully developed (relatively low Reynolds number per the criteria in Dimotakis [30]).

Whether a collision is elastic or inelastic depends upon the magnitude of the mixed-sheet velocity in relation to a critical velocity. The critical velocity is equal to the Taylor-Culick velocity, which is the sheet velocity at the rim of the backward mixed sheet at which surface tension forces are in equilibrium with inertial forces. When the mixed-sheet velocity is greater than the Taylor-Culick velocity, the collision is elastic resulting in the ejection of liquid from the backward mixed sheet (see left photograph in Fig. 2). When the mixed-sheet velocity is less than the Taylor-Culick velocity, the collision is inelastic, and liquid is not ejected from the backward mixed sheet (right photograph of Fig. 2 and result of Eq. 6).

For impinging sheets, it is the equilibrium between surface tension forces and inertial forces **at the rim of the backward mixed sheet** which controls the nature of the collision (inelastic or partially elastic). At the equilibrium condition, a theoretical expression for $COR_{s,c}$ at the critical velocity where the nature of the collision changes from inelastic to elastic was derived [22]:

$$COR_{s,c} = cos\theta + \frac{2R_B}{R_i We_{i,c}} \qquad (8)$$

$R_i$ is the distance from the origin of the expanding single sheets to the impingement zone, $R_B$ is the distance from the origin of the single sheets to the rim of the backward mixed sheet, and $We_{i,c}$ is the Weber number at the impingement zone and at the critical single-sheet velocity ($u_c$) that produces backflow from the mixed sheet. Eq. 8 indicates that it is the second term on the right-hand side that accounts for the "elasticity" of the collision.

Fig. 3 plots $COR_s$ as a function of the velocity of single sheets of various liquids at a nominal impingement angle of 44º [22]. For each liquid, there is a step change in $COR_s$ once the inertial velocity exceeds the Taylor-Culick velocity. $COR_s$ for an inelastic collision is theoretically equal to cosθ and the chart shows that $COR_s$ for inelastic collisions of water is very close to the theoretical. However, for skim milk and 32% aqueous glycerin, the inelastic $COR$ is lower than the theoretical, which suggests that, in addition to the energy associated with the y-component velocity, some of the energy associated with the x-component velocity of the single sheets was dissipated.

Although Eq. 8 is only valid at the critical velocity where surface tension forces are balanced by inertial forces, Fig. 3 indicates that $COR_s$ at higher single-sheet velocities does not significantly vary from that at the critical velocity. Further, $COR_s$ for the partially elastic collisions of the liquids shown in Fig. 3 is essentially identical (allowing for the slight differences in impingement angle from the nominal).

As shown in Fig. 4, the study on $COR_s$ of impinging sheets showed very good agreement between the theoretically calculated $COR_{s,c}$ from Eq. 8 and $COR_{s,c}$ measured from experiments where the mixed-sheet velocity was equal to the Taylor-Culick velocity.

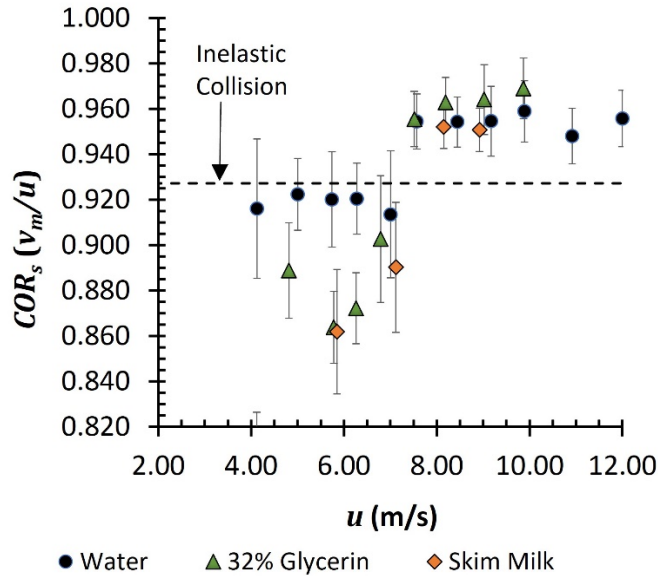

**Fig. 3.** Variation of $COR_s$ with single-sheet velocity for equal impinging sheets of water, 32% glycerin and skim milk. Measured values of θ were 22º for water, 21º for 32% glycerin and 22.5º for skim milk. At the critical velocity for each liquid there is a step or near step change increase in the $COR$ that indicates a partially elastic collision resulting in backflow from the mixed sheet. Figure reproduced from Demyanovich [22].

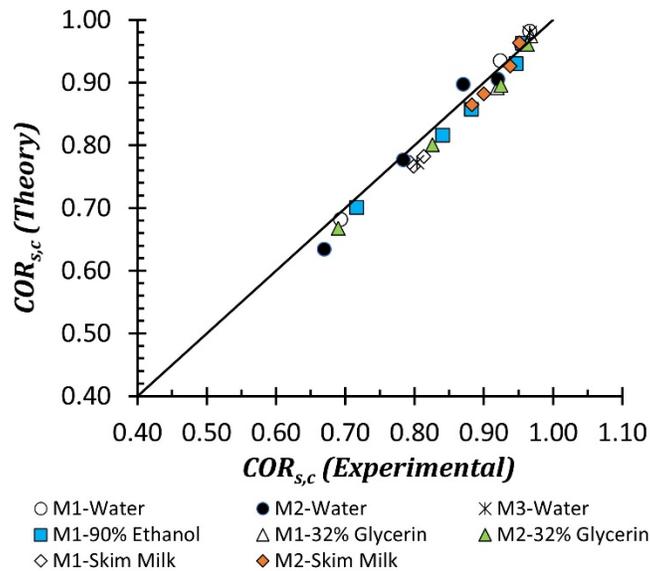

**Fig. 4.** Theoretical critical coefficient of restitution ($COR_{s,c}$) of impinging sheets plotted versus the experimentally determined $COR_{s,c}$ at the critical single-sheet velocity ($u_c$ was determined from plots such as Fig. 3). Three different impinging sheet devices with equivalent orifice diameters of 0.12 mm (M1), 0.19 mm (M2) and 0.27 mm (M3) were investigated. Liquids for the identically impinging sheets were water, 32% glycerin, 90% ethanol and skim milk. Line of equality is included for illustrative purposes. Figure reproduced from Demyanovich [22].

Although the Weber number at the impingement zone is shown in Eq. 8, the equation was derived by assuming that the Weber number at the rim of the backward mixed sheet was equal to 1. The experimental results plotted in Fig. 5 generally verify that the nature of the collision changed from inelastic to partially elastic when the Weber number at the rim of the backward mixed sheet was greater than 1.

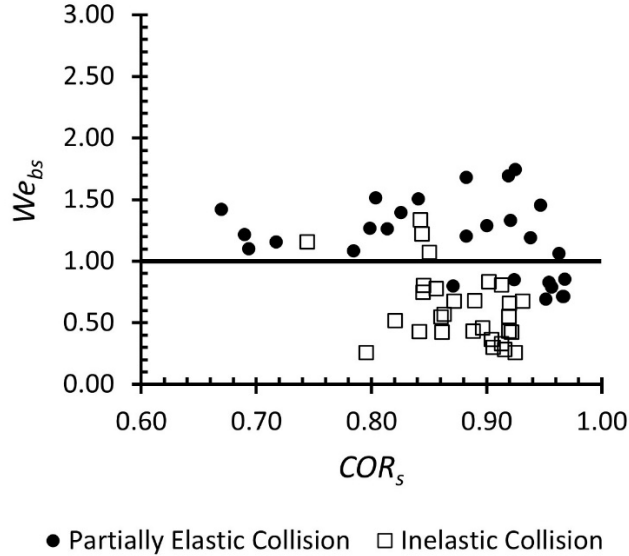

Fig. 5. Plot of the Weber number at the rim of the backward mixed sheet ($We_{bs}$) as a function of the $COR$ of impinging sheets ($COR_s$). For the experimental data, where the measured $COR$ indicated an inelastic collision ($COR_s \approx \cos\theta$), 86% of the data points had a Weber number less than 1. For the experimental data, where the measured $COR$ indicated a partially elastic collision ($\cos\theta < COR_s < 1$), 73% of the data points had a Weber number greater than 1.

**B. Impinging Jets**

*1. Mass and Momentum Balances*

Can the concepts discussed for the geometrically simple system of impinging sheets be applied to impinging jets? Both systems are impinging free streams. There are some important differences, though, which in general complicate the analysis for impinging jets relative to impinging sheets. For impinging jets there is theoretically only one stagnation point in the impingement zone, which is the source point for the production of the liquid sheet. For impinging sheets, however, the impingement zone is a thin line (see Fig. 1b and the right photograph of Fig. 2) and there are stagnation points all along this line.

Further, it has been demonstrated that the velocities of single and mixed sheets, although different, are constant everywhere within the respective sheet. For impinging jets, this is only true for jets which have a flat or uniform profile, such as turbulent jets. According to Anderson et al. [5], a regime where the jets are laminar is "clearly not encountered in rocket engine combustors". Accordingly, the analysis for impinging jets in the present study assumes constant, but not equal, jet and sheet velocities.

Fig. 6a is a schematic depiction of the formation of a liquid sheet from the impingement of two equal jets of liquids with radius, $R$, and velocity, $v_j$. To determine the amount of liquid flowing in various sectors of the sheet, Hasson and Peck [8] considered a section of the jet that was in a plane parallel to the plane of the liquid sheet ("Cross-section through jet" in Fig. 6a). This cross-section through the jet is shown in more detail in Fig. 6b.

A differential mass balance equates input of material flowing in the angular element between $\kappa$ and $\kappa + d\kappa$ of the elliptical cross section in the jet (hatched area in Fig. 6b) to the output in a corresponding angular element in the sheet (hatched area of Fig. 6a):

$$2\left(\rho \frac{p^2}{2} d\kappa (v_j \sin\theta)\right) = \rho h r v_s d\kappa \qquad (9)$$

where $\rho$ is the density of the liquid, $p$ is the distance from the separation point, M, to a point on the ellipse, $h$ is the thickness of the sheet and $r$ is the radial distance in the sheet from the stagnation point, S.

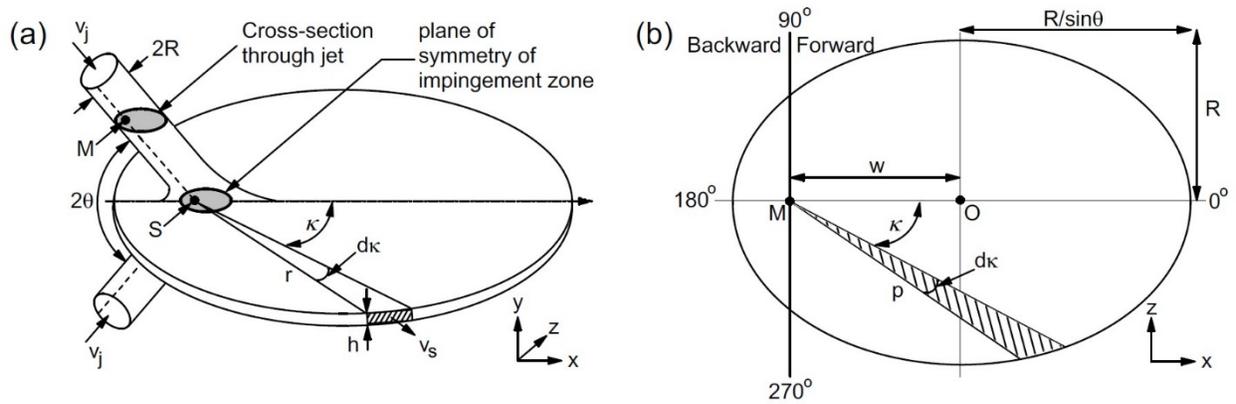

**Fig. 6.** Figure 6a shows the liquid sheet formed by two equal jets impinging at an angle of 90°. Figure 6b illustrates the cross-section through the jet in the x-z plane.  h – sheet thickness; κ - azimuthal angle in sheet and in the cross-section through the jet; r – distance from the stagnation point to the sheet thickness, h; M – separation point in cross-section through jet; p – distance from the separation point M to a point on the ellipse of the cross-section through the jet; R – jet radius; S – stagnation point in the impingement zone; θ – half angle of impingement; v$_j$ – uniform jet velocity; v$_s$ – uniform sheet velocity; w – distance from the center of the ellipse (O) to the separation point, M, or the stagnation point, S.

The left-hand side of Eq. 9 refers to the mass flowing in both jets and the right-hand side refers to the mass flowing in the formed liquid sheet. Simplifying and rearranging yields,

$$\frac{hr}{R^2} = \frac{p^2}{R^2} sin\theta \frac{v_j}{v_s} \qquad (10)$$

Similar to impinging sheets, the coefficient of restitution of the collision of impinging jets is defined as,

$$COR_j = \frac{v_s}{v_j} \qquad (11)$$

yielding,

$$\frac{hr}{R^2} = \frac{p^2}{R^2} \frac{sin\theta}{COR_j} \qquad (12)$$

Previous derivations of the mass balance in most investigations of impinging jets do not include the dissipation of energy at impact and assume that $v_s = v_j$ or $COR_j = 1$.

The momentum balance in the x direction equates the x direction momentum in the two jets with the total x momentum generated in the sheet,

$$2\left(\rho\pi R^2 v_j(v_j cos\theta)\right) = \int_0^{2\pi} \rho hr v_s v_s cos\kappa d\kappa = 2\int_0^{\pi} \rho hr v_s v_s cos\kappa d\kappa \qquad (13)$$

Simplifying and rearranging yields,

$$\pi v_j^2 cos\theta = \int_0^{\pi} \frac{hr}{R^2} v_s^2 cos\kappa d\kappa \qquad (14)$$

Substituting in $COR_j$

$$\frac{\pi cos\theta}{COR_j^2} = \int_0^{\pi} \frac{hr}{R^2} cos\kappa d\kappa \qquad (15)$$

Substituting for $hr/R^2$ from Eq 12 results in

$$\frac{\pi cot\theta}{COR_j} = \int_0^{\pi} \frac{p^2}{R^2} cos\kappa d\kappa \qquad (16)$$

*2. Relationship of $COR_j$ to the location of the stagnation point*

If the elliptical cross section through the jet in Fig. 6a is projected down towards the impingement zone along the separation line, then "M" corresponds to "S", the stagnation point. The elliptical plane of symmetry of the impingement zone is identical to the ellipse created from the section through the jet. An analysis of the sheet thickness parameter of impinging jets requires determining the distance ($w$) from the center of the ellipse to the separation point, M, in Fig. 6b or the stagnation point, S, in Fig. 7. If the collision is 100% elastic, this distance, $b_{HP}$ (Fig. 7), from the stagnation point, $S_{HP}$, is equal to $R\cot\theta$ [8].

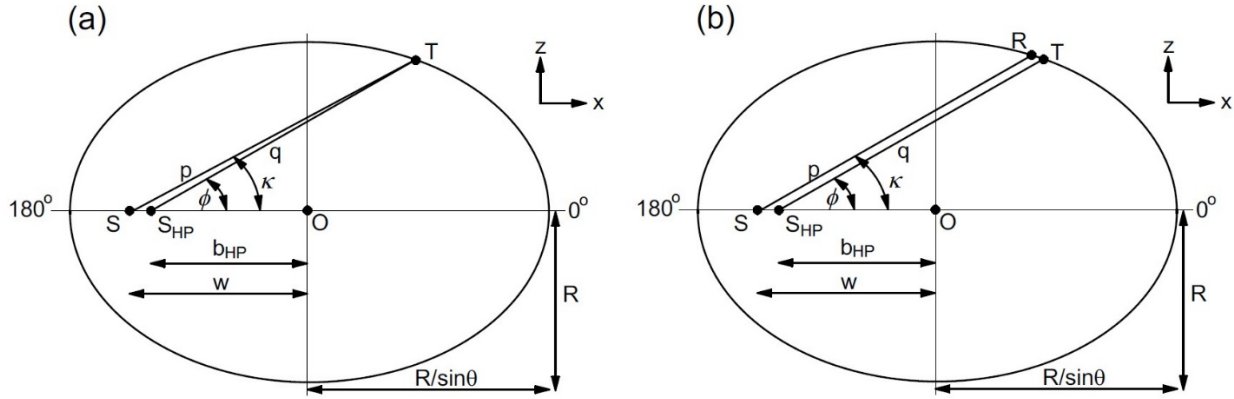

**Fig. 7. Plane of symmetry of the impingement zone illustrating the impact of dissipative collisions on the location of the stagnation point. The plane of symmetry is similar to the cross section through a jet in Fig. 6b. Fig. 7a illustrates that for a specific point on the ellipse, the lengths p and q and the angles $\Phi$ and κ are not equal. Fig. 7b, where $\Phi$ and κ are equal, depicts how the radial line thickness factors for partially elastic collisions will be compared with those for 100% elastic collisions in Section 4 of the present study. $b_{HP}$ – distance, along the ellipse major axis, from the center of the ellipse (O) to the stagnation point, $S_{HP}$, for a 100% elastic collision [8]; κ – azimuthal angle for a partially elastic collision; $\Phi$ – azimuthal angle for a 100% elastic collision; p – length of a vector from S to a point on the ellipse (identical to p in the cross section through the jet); q – length of a vector from $S_{HP}$ to a point on the ellipse; w – distance, along the ellipse major axis, from the center of the ellipse to the stagnation point, S, for a partially elastic collision.**

To determine the location of the stagnation point in the impingement zone or alternatively the separation point in the elliptical cross section, requires mass and momentum balances as well as a relationship between $p$ and $\kappa$. This relationship is given by the polar equation of the ellipse. Note that in the plane of symmetry of the impingement zone, radial lengths from the stagnation point will be referred to as $p$, similar to the jet cross section, but within the liquid sheet beyond the plane of symmetry, $r$ is used to refer to radial distances from the stagnation point.

Referring to Fig. 6b, the polar equation of the ellipse is

$$\left(\frac{p}{R}\cos\kappa - \frac{w}{R}\right)^2 \sin^2\theta + \left(\frac{p}{R}\sin\kappa\right)^2 = 1 \tag{17}$$

Solving equation 17 for $p/R$ yields (derivation provided in the Appendix)

$$\frac{p}{R} = \frac{\frac{w\sin^2\theta\cos\kappa}{R} + \sqrt{\sin^2\kappa + \sin^2\theta\cos^2\kappa - \frac{w^2}{R^2}\sin^2\theta\sin^2\kappa}}{(\sin^2\kappa + \sin^2\theta\cos^2\kappa)} \tag{18}$$

Substituting Eq. 18 into the momentum balance given by Eq. 16 yields,

$$\frac{\pi\cot\theta}{COR_j} = \int_0^\pi \left(\frac{\frac{w\sin^2\theta\cos\kappa}{R} + \sqrt{\sin^2\kappa + \sin^2\theta\cos^2\kappa - \frac{w^2}{R^2}\sin^2\theta\sin^2\kappa}}{(\sin^2\kappa + \sin^2\theta\cos^2\kappa)}\right)^2 \cos\kappa\, d\kappa \tag{19}$$

For 100% elastic collisions, $w = b_{HP}$. For dissipative collisions $COR_j < 1$, and solution of Eq. 19 results in the stagnation point moving towards the rear section of the sheet ($w > b_{HP}$ as shown in Fig. 7). If instead $w < b_{HP}$, then Eq. 19 calculates that $COR_j > 1$, which, based on the definition of $COR_j$ in Eq. 11, is not possible for uniform jet and sheet velocity profiles. This result agrees with observations from a CFD study conducted by Chen et al. [20] where they note that the stagnation point is closer to the rear point of the sheet for a dissipative collision than predicted when assuming a non-dissipative collision.

### 3. Theoretical expression for $COR_j$ based on an equilibrium of surface tension forces and inertial forces

Chen et al. [20] conducted a CFD study on the dynamics and stability of impinging jets. As part of that study, they looked at the stability of the backward liquid sheet (See Fig. 6b). They found that the backward portion of the liquid sheet broke up into droplets (became unstable) as a function of velocity. At low velocities, the flow around the rear point of the liquid sheet was stable. As the velocity increased, the rear liquid sheet became longer and thinner near the rim eventually becoming unstable (forming ligaments and ejecting droplets). The results of the simulation showed that the local Weber number of the sheet was about 1 at the initial unstable location in the backward sheet. These results coincide with the results for impinging sheets as discussed in Section II.A and visually evidenced in Fig. 2.

If it is assumed that the nature of the collision for impinging jets changes in a similar manner to that of impinging sheets, then a theoretical value for $COR_j$ based on the Weber number can be derived. However, as indicated in Fig. 6, the radial lines emanating from the stagnation point to the rim of the backward sheet ($90^o \leq \kappa \leq 270^o$) are of different lengths. For impinging sheets, the distance from the impingement zone to the rim of the backward mixed sheet is quasi constant (see Fig. 2). Thus, the derivation will assume that all locations at the rim of the backward sheet of impinging jets experience a partially elastic collision. This will occur when the sheet Weber number ($We_s$) for the radial line at $\kappa = \pi$ is equal to 1 (where the backward sheet is the thinnest).

From Eq. 12 at $\kappa = \pi$,

$$h_\pi r_\pi = p_\pi^2 \frac{\sin\theta}{COR_j} \qquad (20)$$

Since an equilibrium between surface tension and inertial forces is assumed at the rim of the backward portion of the sheet,

$$\rho h_\pi v_{s,c}^2 = 2\sigma \qquad (21)$$

where $v_{s,c}$ is the velocity of the sheet at which the equilibrium exists ($We_s = 1$) and $\sigma$ is the surface tension. This critical velocity is the Taylor-Culick velocity ($v_{TC}$),

$$v_{TC} = v_{s,c} = \sqrt{\frac{2\sigma}{\rho h_\pi}} \qquad (22)$$

Substituting for $h_\pi$ from Eq. 20 yields

$$v_{s,c}^2 = \frac{2\sigma r_\pi COR_{j,c}}{\rho p_\pi^2 \sin\theta} \qquad (23)$$

where $COR_{j,c}$ is at the critical velocity. Further, since $COR_{j,c} = v_{s,c}/v_{j,c}$ where $v_{j,c}$ is the critical jet velocity that results in the liquid sheet having a velocity equal to the Taylor-Culick velocity, dividing both sides of Eq. 23 by $v_{j,c}^2$ yields

$$COR_{j,c}^2 = \frac{2\sigma r_\pi COR_{j,c}}{\rho v_{j,c}^2 p_\pi^2 \sin\theta} \qquad (24)$$

The Weber number of the jet ($We_{j,c}$) at the critical jet velocity is $\rho d_j v_{j,c}^2/\sigma$ and inserting into Eq. 24 yields ($d_j = 2R$)

$$COR_{j,c} = \frac{4 r_\pi R}{We_{j,c} p_\pi^2 \sin\theta} \qquad (25)$$

From Eq. 18 or Fig. 7 when $\kappa = \pi$,

$$p_\pi = \frac{R}{\sin\theta} - w \qquad (26)$$

Finally,

$$COR_{j,c} = \frac{4r_\pi R}{We_{j,c}\left(\frac{R}{sin\theta} - w\right)^2 sin\theta} \tag{27}$$

Eq. 27 is a theoretical calculation of $COR_{j,c}$ and is only true when the Weber number of the sheet at $\kappa = \pi$ is equal to 1. Further, similar to impinging sheets, at velocities greater than $v_{j,c}$, it will be assumed that $COR_j$ is equal to $COR_{j,c}$ (see Fig. 3).

Comparing $COR_{j,c}$ calculated from Eq. 27 with $COR_{s,c}$ calculated from Eq. 8 indicates similarities between impinging jets and impinging sheets. Both expressions are functions of the critical Weber number of the system and both are functions of characteristic length scales ($R$ and $r_\pi$ for impinging jets and $R_i$ and $R_B$ for impinging sheets). However, the expression for impinging jets indicates that the location of the stagnation point is an important factor.

If the collision of impinging jets is 100% elastic, then $COR_j = 1$, $w = Rcot\theta$, and $r_\pi$ is calculated from Eq 27 as

$$r_\pi = \frac{We_{j,c}\left(\frac{R}{sin\theta} - \frac{Rcos\theta}{sin\theta}\right)^2 sin\theta}{4R} = \frac{We_{j,c}R(1-cos\theta)^2}{4sin\theta} \tag{28}$$

Multiplying by $sin^4\theta/(1-cos^2\theta)^2$ yields,

$$r_\pi = \frac{We_{j,c}Rsin^3\theta(1-cos\theta)^2}{4(1+cos\theta)^2(1-cos\theta)^2} = \frac{We_{j,c}Rsin^3\theta}{4(1+cos\theta)^2} \tag{29}$$

Eq. 29 agrees with Bremond and Villermaux [32] who calculated the rear extension of the sheet at a sheet azimuthal angle equal to π for impinging jets, which had uniform velocity profiles, and for a collision that was non-dissipative (i.e. 100% elastic collision).

Eq. 8 for impinging sheets and Eq. 27 for impinging jets theoretically demonstrate the relationship between the $COR$, the impingement angle, the pre-impinging stream velocity, the physical properties of the impinging liquids, and for impinging jets the stagnation point. However, these equations are difficult to employ for the calculation of $COR_{s,c}$ or $COR_{j,c}$ because they require knowledge of the critical single-sheet velocity or critical jet velocity in order to calculate the critical Weber number (single-sheet $We$ at the impingement zone or jet $We$). This velocity changes depending upon the impingement angle, equivalent orifice diameter, and the physical properties of the liquid(s). Further, for impinging jets, the location of the stagnation point and the distance of the rear extension ($r_\pi$) must be known. For impinging sheets, the length of the backward mixed sheet ($R_B$) is also required but for many liquids, with the exception of skim milk, $R_B$ was found to be less than 2% greater than $R_i$ and thus it could be assumed that $R_B/R_i \approx 1$ in Eq. 8 [22].

Based on the results of experiments, a simpler correlation for the $COR$ of impinging sheets was provided in Demyanovich [22]. A similar correlation for the $COR$ of impinging jets is provided in Section III.C.

### III. Determination of the $COR$ of impinging jets from experimental data and CFD results

Three sets of data from the literature were used to estimate the $COR$ for impinging jets. Two sets of data included experimentally measured velocities while the third set provided the calculated location (from a CFD study) of the stagnation point for dissipative collisions.

**A. Experimentally measured velocities**

*1. Experimental data from Dombrowski and Hooper [10]*

Dombrowski and Hooper [10] studied the velocity at the center line of the sheet (azimuthal angle = 0º in Fig. 6b and 7). Using high-speed cinematography, they measured the distances traveled by surface irregularities and waves. Sheet velocities were measured for both turbulent and laminar jets. The jets were deliberately made turbulent (resulting in uniform velocity profiles) by inserting wires into the flared entry of the tubes producing the jets. Since the present study only considers jets with uniform velocity profiles, only the velocity data for the turbulent jets is included in the analysis.

Table 1 lists the measured velocities for the impingement of turbulent jets. While actual values for the mean jet velocities were provided, the measured sheet velocities were only provided in graphical form and there is some error in reading these values (of the order of a few percent).

$COR_j$ was calculated from Eq. 11. The nature of the collision was determined by comparing the calculated $COR_j$ with cosθ, and if $COR_j$ was significantly greater, the collision was deemed to be partially elastic. The last column of Table 1 indicates if the experimental data was used to create summary statistics. Only the calculated $COR$ at θ = 55º and at $v_j$ = 11.6 m/s, which indicated an inelastic collision, was considered unreliable. The reason is that at the lower velocity of 7.3 m/s and the higher velocity of 16.0 m/s, the collisions are clearly partially elastic. As shown in Fig. 3 for impinging sheets, once a collision becomes partially elastic, it remains so when the velocity is increased. Further, based on experience with impinging sheets, 2θ = 110º is a high collision angle (with respect to the $COR$) and inelastic collisions only occur at very low velocities.

**TABLE 1.** Measurements of velocities for impinging, equal turbulent jets and at the center line ($\kappa = 0$) of the liquid sheet formed from the collision [10]. L – length of the free jet from the exit of the tube to impingement; Part. – partially. $d_j$ = 0.5mm, $L/d_j$ = 4.

| θ | $v_j$ | $v_s$ (κ = 0°) | Calc. $COR_j$ | cosθ | Collision Type | Included in Table 2 Statistics |
|---|---|---|---|---|---|---|
| 25 | 7.3 | 6.67 | 0.91 | 0.91 | Inelastic | Yes |
| 25 | 11.6 | 10.56 | 0.91 | 0.91 | Inelastic | Yes |
| 25 | 16.0 | 14.40 | 0.90 | 0.91 | Inelastic | Yes |
| 25 | 19.5 | 16.33 | 0.84 | 0.91 | Inelastic | Yes |
| 40 | 7.3 | 5.83 | 0.80 | 0.77 | Inelastic | Yes |
| 40 | 11.6 | 8.79 | 0.76 | 0.77 | Inelastic | Yes |
| 40 | 16.0 | 12.50 | 0.78 | 0.77 | Inelastic | Yes |
| 40 | 19.5 | 15.50 | 0.79 | 0.77 | Inelastic | Yes |
| 55 | 7.3 | 5.08 | 0.70 | 0.57 | Part. Elastic | Yes |
| 55 | 11.6 | 6.79 | 0.59 | 0.57 | ?? | No |
| 55 | 16.0 | 11.92 | 0.74 | 0.57 | Part. Elastic | Yes |
| 55 | 19.5 | 15.00 | 0.77 | 0.57 | Part. Elastic | Yes |
| 70 | 7.3 | 3.92 | 0.54 | 0.34 | Part. Elastic | Yes |
| 70 | 11.6 | 7.08 | 0.61 | 0.34 | Part. Elastic | Yes |
| 70 | 16.0 | 11.08 | 0.69 | 0.34 | Part. Elastic | Yes |
| 70 | 19.5 | 14.13 | 0.72 | 0.34 | Part. Elastic | Yes |

Table 2 lists the summary statistics of the individual data measurements listed in Table 1. The resulting mean $COR_j$ (column 2) is compared with $COR_s$ (columns 4 and 5) at the same impingement angle. No value for $COR_s$ at $\theta = 70°$ is listed in column 4 because the measured $COR$ data for impinging sheets only encompasses $20° \leq \theta \leq 60°$. Matches between the $COR$ values of impinging jets and impinging sheets are highlighted in each row.

**TABLE 2.** Summary of the data listed in Table 1 and comparison of calculated $COR_j$ with $COR_s$ at the same half angle of impingement. Matches between jet and sheet $COR$ values are highlighted in each row. % S.D. – standard deviation divided by the mean $COR$ and reported as a percentage.

| θ (°) | Mean $COR_j$ (imp. jets) | % S.D. of $COR_j$ | Partially elastic $COR_s$ (imp. sheets) | Inelastic $COR$ (cosθ) |
|---|---|---|---|---|
| 25 | 0.89 | 4.0% | 0.95 | 0.91 |
| 40 | 0.78 | 2.4% | 0.86 | 0.77 |
| 55 | 0.73 | 4.7% | 0.72 | 0.57 |
| 70 | 0.64 | 13.2% | N/A | 0.34 |

For impingement angles (2θ) of 50º and 80º the collisions were inelastic, which is not surprising since 2θ is relatively low for the velocities investigated and because water has a high surface tension and thus relatively high Taylor-Culick velocity. At high impingement angles for impinging sheets, partially elastic collisions occur at relatively low velocities. This also seems to be the case for impinging jets at impingement angles of 110º and 140º. Further, the measured $COR_j$ at 2θ = 110º matches well with $COR_s$ for a partially elastic collision (no comparison can be made at 2θ = 140º because this impingement angle is outside the scope of the experiments designed to measure the $COR$ of impinging sheets).

## 2. *Experimental data from Choo and Kang [4]*

Choo and Kang [4] measured the velocity of liquid elements produced by the disintegration of the liquid sheet formed by the impingement of two equal jets of water. They defined liquid elements as the sheet and the ligaments emanating from the edge of the sheet when it breaks down. While they only measured the velocity of the ligaments, the inference is that this velocity applies to the liquid sheet as well since both are considered liquid elements. This inference is supported by the fact that while discussing available literature references on the velocity characteristics of "liquid elements", the study by Dombrowski and Hooper [10] which measured sheet velocities, is cited.

The experimental data were taken at "high speed" (mean jet velocities of 6.76 to 12.38 m/s) and "low speed" (mean jet velocity of 3.35 m/s). Although the velocity profiles in the jets were fully developed, parabolic velocity profiles within tubes are known to relax to a more uniform profile when the free jet is formed [21,32]. In an earlier study by Choo and Kang [15] on the velocity distribution in a liquid sheet that was formed by the impingement of two low-speed jets, it was found that the sheet velocity is a function of azimuthal angle. However, in that study, the authors also state that the sheet velocities can be approximated as the mean jet velocity in some conditions such as a short orifice nozzle and much higher jet velocities. For low-speed jets the central region (around the center line of the liquid sheet) has a significantly higher velocity than the mean jet velocity.

For the "high-speed" study [4], all measured ligament velocities are less than the mean jet velocity and within the central region, the measured ligament velocities are approximately constant. Thus, it will be assumed that the "high-speed" velocity data are applicable to the present study and since the turbulent jet velocity data from Dombrowski and Hooper [10] were only taken at the center line, only velocity measurements at the center line of the Choo and Kang [4] data will be analyzed.

Table 3 lists the measured velocities for the impingement of high-velocity jets. Actual values for the mean jet velocities were provided, but measured ligament velocities were only provided in graphical form, leading to some error in reading these values (of the order of a few percent). The mean $COR_j$ for the data at θ = 80° in Table 3 is 0.87 with a standard deviation of 3.0% of the mean.

A comparison of columns 4 and 5 indicates very good agreement between the $COR$ for the impinging jets and the $COR$ for partially elastic collisions of impinging sheets (the mean value of 0.87 at θ = 80° for impinging jets is used for this comparison).

**TABLE 3.** Measurements of velocities in impinging equal jets and the liquid sheet formed from the collision [4]. L – length of the free jet from the exit of the tube to impingement; Part. – partially.  $d_j$ = 1 mm, $L/d_o$ = 10.

| θ | $v_j$ | $v_s$ (κ = 0°) | Calc. $COR_j$ (imp. jets) | Partially elastic $COR_s$ (imp. sheets) | cosθ | Collision Type |
|---|---|---|---|---|---|---|
| 80 | 6.76 | 5.7 | 0.84 | 0.86 | 0.77 | Part. Elastic |
| 80 | 9.50 | 8.5 | 0.89 | 0.86 | 0.77 | Part. Elastic |
| 80 | 12.38 | 10.8 | 0.87 | 0.86 | 0.77 | Part. Elastic |
| 100 | 6.76 | 5.1 | 0.76 | 0.77 | 0.64 | Part. Elastic |
| 120 | 6.76 | 4.4 | 0.65 | 0.66 | 0.5 | Part. Elastic |

**B. Simulation of stagnation point location**

Chen et al. [20] conducted a CFD study on the dynamics and stability of impinging jets. One of the findings from the simulations was that an assumption of a 100% elastic, non-dissipative impact of impinging jets is "not true for real case". They found that the location of the stagnation point varied from that calculated assuming a 100% elastic collision. The stagnation point was shifted towards the backward sheet.

The location of the stagnation point $(w/R)$ determined in the CFD study is listed in Table 4. Eq. 19 was numerically integrated to determine $COR_j$ from these locations of the stagnation point at the three impingement angles. The agreement between the numerically calculated value of $COR_j$ and $COR_s$ is very good except at the highest half angle of impingement.

**TABLE 4.** Calculation of $COR_j$ from numerical integration of Eq. 19 using stagnation point locations ($w/R$) for dissipative collisions as determined from the CFD study by Chen et al. [20].

| θ (°) | $w/R$ | $COR_j$ calc. from Eq. 19 | Partially elastic $COR_s$ (imp. sheets) |
|---|---|---|---|
| 27.5 | 2.09 | 0.95 | 0.94 |
| 44.5 | 1.34 | 0.82 | 0.82 |
| 59 | 0.86 | 0.73 | 0.67 |

## C. Correlation for the $COR$ of Impinging Jets

Fig. 8 plots $COR_j$ determined from the velocity data and the stagnation-point location data. As noted earlier, $COR_j$ calculated from the velocity data is in good agreement with $COR_s$ for impinging sheets. Further, $COR_j$ calculated from the stagnation-point location data agrees well with $COR_s$ except at 2θ = 118°, which is at the highest 2θ at which $COR_s$ was experimentally investigated.

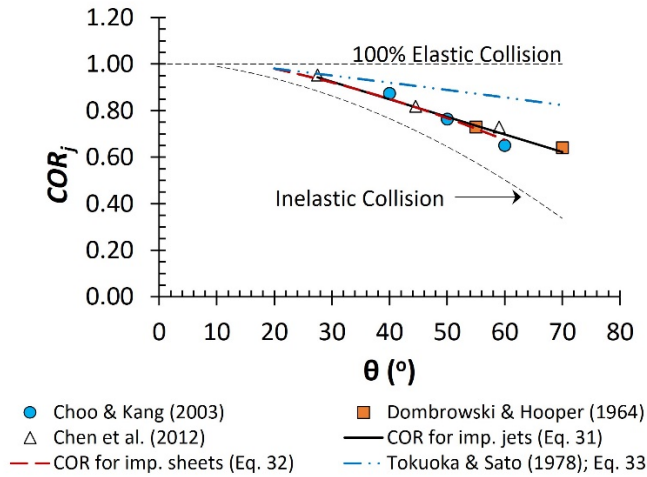

**Fig. 8.** Plot of $COR_j$, determined from velocity measurements [4,10] and from the results of a CFD study on the location of the stagnation point [20], versus θ. —— impinging-jet correlation (Eq. 31); – – impinging-sheet correlation (Eq. 32); —··— correlation from Tokuoka and Sato [18] (Eq. 33).

For impinging sheets, the theoretical expression for $COR_{s,c}$ (Eq. 8) suggests that an equation of fit of $COR_s$ with θ could be based on cosθ (for an inelastic collision) plus the difference between the partially elastic curve and the inelastic curve. Thus, the equation of fit for the partially elastic $COR$ of impinging sheets was determined to be

$$COR_s = cos\theta + 0.003268(\theta) - 0.03783 \tag{30}$$

where θ is in degrees [22]. The regression coefficient for this fit is 0.977.

For impinging jets, however, the theoretical expression given by Eq. 27 does not readily provide a similar clue as to the expected fit for the experimental and CFD data. As a result, the calculated values of the partially elastic $COR$ from the experimental velocity data and CFD results were plotted against θ and fit to 95 different regression models, including many three and four-parameter fit models. The top-rated model was the following simple regression

$$COR_j = cos\theta + 0.1085/cot\theta \quad (31)$$

where θ is in radians (27.5° < θ < 70°) and the regression coefficient is 0.962. While the form of the regression equation is not surprising (given the results for impinging sheets), it is interesting that the elastic contribution to the $COR$ is a single term including the cotθ, which is present in the momentum balance (Eq. 16) of impinging jets. This finding led to a reassessment of the expression for impinging sheets in that perhaps the two terms that calculate the contribution of the elasticity of the collision to $COR_s$ in Eq. 30 might be collapsed into one term. Of course, since the $COR$ for impinging sheets and impinging jets more or less coincide, an expression such as Eq. 31 would be expected to apply.

A similar expression, then, for the fit of the experimental measurements of the $COR$ for impinging sheets is

$$COR_s = cos\theta + 0.1021/cot\theta \quad (32)$$

where θ is in radians and the regression coefficient is 0.972. Although the fit is not quite as good as Eq. 30, Eq. 32 has the advantage that the elasticity of the collision of impinging sheets is represented by one term as it is in the theoretical equation ($\frac{2R_B}{R_i We_c}$ in Eq. 8). Interestingly, at θ = 70°, which is outside the range of the experimental data for impinging sheets, Eq. 30 calculates $COR_s$ equal to 0.53 whereas Eq. 32 calculates 0.62. $COR_s$ calculated from Eq. 32 is much closer to $COR_j$, which is equal to 0.64 when θ = 70°.

Earlier it was mentioned that Tokuoka and Sato [18] provided a correlation for estimating the velocity of the liquid sheet based on an experimental value at 2θ = 180°. The correlation is

$$\Psi = \left(sin^2\theta \Psi_{180°}^2 + cos^2\theta\right)^{0.5} \quad (33)$$

where $\Psi$ is termed the coefficient of effective velocity. $\Psi_{180°}$ is the value of $\Psi$ for an impingement angle of 180°, which was stated as having been shown experimentally to be 0.8. However, it is not clear how Eq. 33 was developed in the absence of any further data.

Presumably, $\Psi = COR_j$. Calculations from this correlation are shown in Fig. 8. Eq. 33 predicts significantly higher values of $COR_j$ when compared with impinging sheets and the velocity measurement data sets and CFD results discussed earlier for impinging jets.

The chart in Fig. 8 indicates that for $20° \leq \theta \leq 60°$ both the impinging-jet expression (Eq. 31) and impinging-sheet expression (Eq. 30 or 32) yield similar values of $COR_j$. If θ > 60°, then the correlation for impinging jets should be used as it appears to fit the very high impingement angles better and the $COR$ correlation for impinging sheets is only considered to be valid up to θ = 60°.

### IV. Effect of dissipative collisions on thickness distribution in the liquid sheet

The liquid sheet formed by the impingement of two jets expands ultimately breaking up into droplets. The distribution of drop sizes detaching from the rim of the liquid sheet depends upon the thickness and the orientation of the rim [33]. Drop size and drop-size distributions are important factors in the vaporization rate, which, as noted in the Introduction, is often the rate-controlling step of combustion within liquid rocket engines [2,3].

Several equations have been proposed for the "thickness parameter" of the liquid sheet, which is defined as $hr/R^2$ where $R$ is the jet radius, $h$ is the thickness of the sheet and $r$ is the distance from the stagnation point to the thickness, $h$ [6-9]. These equations generally have the form

$$\frac{hr}{R^2} = f(\theta, \kappa) \tag{34}$$

The equation provided by Hasson and Peck [8] is

$$\frac{hr}{R^2} = \frac{\sin^3 \theta}{(1 - \cos\Phi \cos\theta)^2} \tag{35}$$

where $\Phi$ is the angle shown in Fig. 7 as measured at the stagnation point located at S$_{HP}$ ($COR_j = 1$ and $\kappa = \Phi$). The correlation provided by Ibrahim and Przekwas [6] is

$$\frac{hr}{R^2} = \frac{\beta}{(e^\beta - 1)} e^{\beta(1 - \frac{\Phi}{\pi})} \tag{36}$$

where $\Phi$ is as defined for Eq. 35 and β is determined by the impingement angle and is numerically calculated from

$$\cos\theta = \left[\frac{(e^\beta + 1)}{(e^\beta - 1)}\right] \frac{1}{1 + \left(\frac{\pi}{\beta}\right)^2} \tag{37}$$

These correlations for the thickness parameter assume a non-dissipative or 100% elastic collision. Thus, it is of interest to determine the impact of dissipative collisions on the thickness parameter. For partially elastic collisions, the thickness parameter is derived by substituting Eq. 18 into Eq. 12 yielding

$$\frac{hr}{R^2} = \left( \frac{\frac{w \sin^2\theta \cos\kappa}{R} + \sqrt{\sin^2\kappa + \sin^2\theta \cos^2\kappa - \frac{w^2}{R^2}\sin^2\theta \sin^2\kappa}}{(\sin^2\kappa + \sin^2\theta \cos^2\kappa)} \right)^2 \frac{\sin\theta}{COR_j} \quad (38)$$

$w/R$ is determined from Eq. 19 with $COR_j$ calculated from Eq. 31. If $COR_j = 1$, then $w/R = b_{HP}/R = \cot\theta$ and Eq. 38 reduces to Eq. 35.

In Fig. 9, the thickness parameter predicted assuming a dissipative collision is compared (per Fig. 7b) with the thickness parameters derived assuming a non-dissipative collision. Partially elastic collisions increase the thickness parameter in the core section of the forward part of the sheet and decrease the thickness parameter outside of this section. Where this transition occurs is a function of θ. For impinging-jet injectors used in rocket engines, where 2θ typically is 60º, this transition occurs at an azimuthal angle of approximately 30º and by symmetry 330º (or -30º) which increases the amount of mass flowing in this central core of the sheet. However, the amount of backspray towards the impinging-jet injector will decrease. This is expected as Figs. 6b and 7 show that as the stagnation point moves closer to the rear point of the sheet, the amount of flow in the forward sheet is increased and in the backward sheet is decreased.

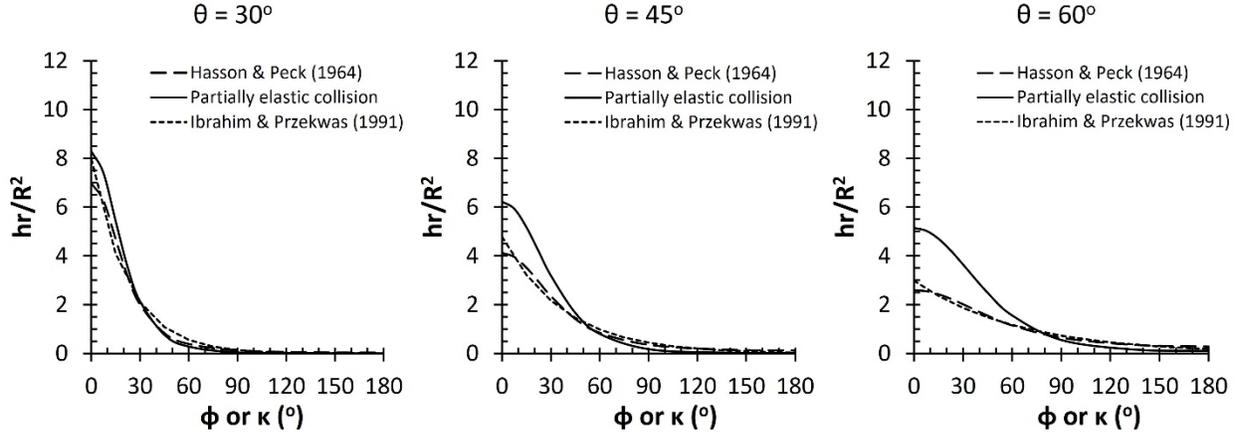

**Fig. 9. Calculated thickness parameter versus azimuthal angle for the impingement of two jets, each having a uniform velocity profile at impingement. Comparisons are made based on Fig. 7b. $COR = 1$ for the correlations of Hasson and Peck [8] and Ibrahim and Przekwas [6]. For partially elastic collisions, $\cos\theta < COR < 1$.**

Hasson and Peck [8] and Ibrahim and Przekwas [6] compared their correlations with experimental data from Taylor [34]. However, no such comparison with partially elastic collisions is made here for two reasons. First, the velocity of the jets in Taylor's experiments was very low (pressure head was less than 200 cm of water (0.2 bar)) and flow straighteners were used to reduce/eliminate turbulence in the jets. Second, Taylor used sharp-edged orifices in his study but did not provide the coefficient of contraction. As implied by Choo and Kang [33], both Hasson and Peck [8] and Ibrahim and Przekwas [6] chose coefficients of contraction (0.64 and 0.72, respectively) that yielded good agreement between their predictions and Taylor's results.

With regards to more recent measurements [33,35], these studies were carried out at low jet velocities and non-uniform jet and sheet velocity profiles. As mentioned several times, the work presented here is for uniform jet and sheet velocity profiles.

The calculation of the thickness parameter for partially elastic collisions was somewhat cumbersome requiring three equations with one equation involving an integral that must be numerically integrated (Eq. 19). This process can be simplified by recognizing that the location of the stagnation point, $w/R$, correlates very well with the parameter $\cot\theta/COR_j$ as shown in Fig. 10. The equation of fit is

$$\frac{w}{R} = 1.03 \frac{\cot\theta}{COR_j} \tag{39}$$

with a correlation coefficient > 0.99. For the calculation of the thickness parameter from Eq. 38, this simplified equation can be used to replace Eq. 19.

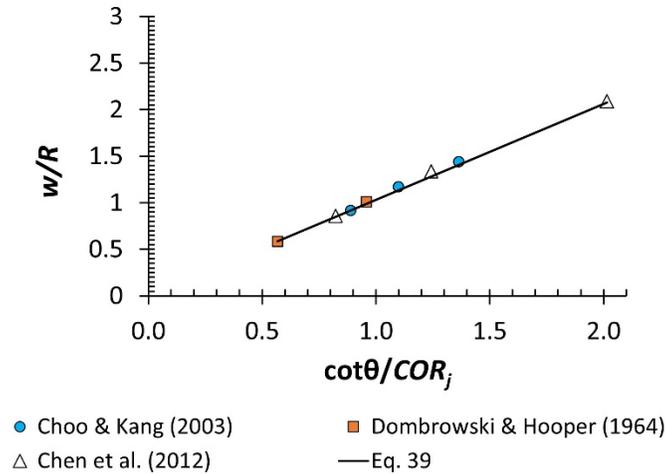

**Fig. 10. Plot of $w/R$ versus $\cot\theta/COR_j$. The equation of fit is given by Eq. 39 and can be used as a substitute for Eq. 19, which requires numerical integration to solve.**

## V. Conclusions

For impingement angles between 40 and 110°, the coefficient of restitution ($COR$) of the collision of impinging jets ($COR_j$) is equal to within 1 to 2% of the $COR$ of the collision of impinging sheets. Calculations from available literature data on jet and sheet velocities indicate that impinging jets undergo inelastic and partially elastic collisions similar to impinging sheets. A simple empirical correlation (Eq. 31) was developed that relates $COR_j$ to the half angle of impingement.

For dissipative collisions, the stagnation point shifts, along the major axis of the elliptical plane of symmetry of the impingement zone, towards the backward section of the sheet. The location of the stagnation point is related to $COR_j$ and can be determined from numerical integration of the momentum balance (Eq. 19) or from a simple empirical relationship with $\cot\theta/COR_j$ (Eq. 39).

A theoretical expression is provided for the impact of dissipative collisions on the thickness distribution in the formed liquid sheet (Eq. 38). For impingement at oblique angles, the thickness of the core region of the forward sheet is thicker and the thickness of the region outside the core is thinner, relative to the non-dissipative assumption.

This investigation is limited to uniform velocity profiles in both the impinging jets and the formed sheet, which is the condition believed to prevail in rocket engines. Further study is recommended for the case where jet and sheet velocities have non-uniform profiles and where local sheet velocities are greater than the mean jet velocity.

## Appendix

The polar equation (Eq. 17) for the ellipse shown in Fig. 6b and 7 is:

$$\left(\frac{p}{R}\cos\kappa - \frac{w}{R}\right)^2 \sin^2\theta + \left(\frac{p}{R}\sin\kappa\right)^2 = 1 \tag{A-1}$$

Expanding this expression yields

$$\frac{p^2}{R^2}\sin^2\kappa + \left(\frac{p^2}{R^2}\cos^2\kappa + \frac{w^2}{R^2} - 2\frac{pw\cos\kappa}{R^2}\right)\sin^2\theta = 1 \tag{A-2}$$

and,

$$(\sin^2\kappa + \cos^2\kappa \sin^2\theta)\frac{p^2}{R^2} - \left(2\frac{w\cos\kappa}{R}\sin^2\theta\right)\frac{p}{R} + \left(\frac{w^2}{R^2}\sin^2\theta - 1\right) = 0 \tag{A-3}$$

The solution for $p/R$ in Eq. A-3 is of the form,

$$\frac{p}{R} = \frac{-B \pm \sqrt{B^2 - 4AC}}{2A} \tag{A-4}$$

where,

$$A = sin^2\kappa + cos^2\kappa sin^2\theta \tag{A-5}$$

$$B = -\frac{2w cos\kappa sin^2\theta}{R} \tag{A-6}$$

$$C = \frac{w^2}{R^2}sin^2\theta - 1 \tag{A-7}$$

and

$$B^2 - 4AC = \frac{4w^2 cos^2\kappa sin^4\theta}{R^2} - 4(sin^2\kappa + cos^2\kappa sin^2\theta)(\frac{w^2}{R^2}sin^2\theta - 1) \tag{A-8}$$

Expanding Eq. A-8 yields,

$$B^2 - 4AC = \frac{4w^2 cos^2\kappa sin^4\theta}{R^2} - 4\frac{w^2}{R^2}sin^2\theta sin^2\kappa + 4sin^2\kappa - 4\frac{w^2}{R^2}cos^2\kappa sin^4\theta + 4cos^2\kappa sin^2\theta \tag{A-9}$$

The first and fourth terms on the right-hand side in Eq. A-9 cancel out, leading to

$$B^2 - 4AC = -4\frac{w^2}{R^2}sin^2\theta sin^2\kappa + 4sin^2\kappa + 4cos^2\kappa sin^2\theta \tag{A-10}$$

Inserting Eqs. A-5, A-6 and A-10 into Eq. A-4 yields

$$\frac{p}{R} = \frac{\frac{2w cos\kappa sin^2\theta}{R} \pm \sqrt{-4\frac{w^2}{R^2}sin^2\theta sin^2\kappa + 4sin^2\kappa + 4cos^2\kappa sin^2\theta}}{2(sin^2\kappa + cos^2\kappa sin^2\theta)} \tag{A-11}$$

Simplifying,

$$\frac{p}{R} = \frac{\frac{w sin^2\theta cos\kappa}{R} + \sqrt{sin^2\kappa + sin^2\theta cos^2\kappa - \frac{w^2}{R^2}sin^2\theta sin^2\kappa}}{(sin^2\kappa + sin^2\theta cos^2\kappa)} \tag{A-12}$$

where only the first root (+ √) in the numerator of Eq. A-11 is retained. As shown in Table A.1, the second root solution (- √) results in absolute values of $p/R$ that are reversed from the first root solution (i.e. the absolute value of the length, $p/R$, at κ = 0° is calculated as the length at κ = 180° (see Fig. 6b)).

If the collision is 100% elastic, then $w/R = cot\theta$, and Eq. A-12 yields the solution provided in Hasson and Peck [8],

$$\frac{p}{R} = \frac{cos\theta cos\kappa sin\theta + sin\theta}{(sin^2\kappa + sin^2\theta cos^2\kappa)} = \frac{sin\theta}{(1 - cos\theta cos\kappa)} \tag{A-13}$$

**TABLE A.1.** Comparison of the first and second root solutions of $\boldsymbol{p/R}$ for Eq. A-11 when $\theta = 30°$.

| $\theta$ (°) | w/R | $\kappa$(°) | $p/R = \frac{-B+\sqrt{B^2-4AC}}{2A}$ | $p/R = \frac{-B-\sqrt{B^2-4AC}}{2A}$ |
|---|---|---|---|---|
| 30 | 1.922 | 0 | 3.922 | -0.0785 |
| 30 | 1.922 | 30 | 1.990 | -0.0884 |
| 30 | 1.922 | 60 | 0.7223 | -0.1311 |
| 30 | 1.922 | 90 | 0.2774 | -0.2774 |
| 30 | 1.922 | 120 | 0.1311 | -0.7223 |
| 30 | 1.922 | 150 | 0.0884 | -1.9902 |
| 30 | 1.922 | 180 | 0.0785 | -3.9215 |

[8]   Hasson, D., and Peck, R. E., "Thickness Distribution in a Sheet Formed by Impinging Jets," *AIChE Journal*, Vol. 10, No. 5, 1964, pp. 752–754.

      https://doi.org/10.1002/aic.690100533

[9]   Ranz, W. E., "Some Experiments on the Dynamics of Liquid Films," *Journal of Applied Physics*, Vol. 30, No. 12, 1959, pp. 1950–1955.

      https://doi.org/10.1063/1.1735095

[10]  Dombrowski, N. D., and Hooper, P. C., "A Study of the Sprays Formed by Impinging Jets in Laminar and Turbulent Flow," *Journal of Fluid Mechanics*, Vol. 18, No. 3, 1964, pp. 392–400.

      https://doi.org/10.1017/S0022112064000295

[11]  Li, R., and Ashgriz, N., "Characteristics of Liquid Sheets Formed by Two Impinging Jets," *Physics of Fluids*, Vol. 18, No. 8, 2006.

      https://doi.org/10.1063/1.2338064

[12]  Anderson, W., Ryan, H., Pal, S., and Santoro, R., "Fundamental Studies of Impinging Liquid Jets," AIAA Paper 92-0458, 1992.

      https://doi.org/10.2514/6.1992-458

[13]  Foster, H. H., and Heidmann, M. F., "Spatial Characteristics of Water Spray Formed by Two Impinging Jets at Several Jet Velocities in Quiescent Air," NASA-TN-D-301, 1960.

      https://ntrs.nasa.gov/api/citations/19980235515/downloads/19980235515.pdf

[14]  Ryan, H. M., Anderson, W. E., Pal, S., and Santoro, R. J., "Atomization Characteristics of Impinging Liquid Jets," *Journal of Propulsion and Power*, Vol. 11, No. 1, 2012, pp. 135–145.

      https://doi.org/10.2514/3.23851

[15]  Choo, Y. J., and Kang, B. S., "The Velocity Distribution of the Liquid Sheet Formed by Two Low-Speed Impinging Jets," *Physics of Fluids*, Vol. 14, No. 2, 2002, pp. 622–627.

      https://doi.org/10.1063/1.1429250

[16]  Lai, W. H., Huang, W., and Jiang, T. L., "Characteristic study on the like-doublet impinging jets atomization," *Atomization and Sprays*, Vol. 9, No. 3, 1999, pp. 277–289.

      https://doi.org/10.1615/ATOMIZSPR.V9.I3.30

[17]  Heidmann, M., Priem, R., and Humphrey, J. C., "A Study of Sprays Formed by Two Impinging Jets," NASA-TN-3835, 1957.

      https://ntrs.nasa.gov/api/citations/19930084568/downloads/19930084568.pdf

[18]  Tokuoka, N., and Sato, G. T., "The Study of a Liquid Atomization by Impinging of Two Jets," *Bulletin of the JSME*, Vol. 21, No. 155, 1978, pp. 885–892.
30

[31] Demyanovich, R. J., "On the Impingement of Free, Thin Sheets of Liquids - A Photographic Study of the Impingement Zone," *AIP Advances*, Vol. 11, No. 1, 2021.
https://doi.org/10.1063/5.0040336

[32] Bremond, N., and Villermaux, E., "Atomization by Jet Impact," *Journal of Fluid Mechanics*, Vol. 549, 2006, pp. 273–306.
https://doi.org/10.1017/S0022112005007962

[33] Choo, Y. J., and Kang, B. S., "Parametric Study on Impinging-Jet Liquid Sheet Thickness Distribution Using an Interferometric Method," *Experiments in Fluids*, Vol. 31, No. 1, 2001, pp. 56–62.
https://doi.org/10.1007/S003480000258

[34] Taylor, G., "Formation of Thin Flat Sheets of Water," *RSPSA*, Vol. 259, No. 1296, 1960, pp. 1–17.
https://doi.org/10.1098/RSPA.1960.0207

[35] Shen, Y. B., and Poulikakos, D., "Thickness Variation of a Liquid Sheet Formed by Two Impinging Jets Using Holographic Interferometry," *Journal of Fluids Engineering*, Vol. 120, No. 3, 1998, pp. 482–487.
https://doi.org/10.1115/1.2820688